\documentclass[conference]{IEEEtran}
\usepackage{ifpdf}
\usepackage{cite}

\newtheorem{definition}{\bf Definition}

\ifCLASSINFOpdf
\else
\fi

\usepackage[cmex10]{amsmath}
\usepackage{algorithmic}
\usepackage{array}
\usepackage{mdwmath}
\usepackage{mdwtab}
\usepackage{eqparbox}
\usepackage[tight,footnotesize]{subfigure}
\usepackage{fixltx2e}
\usepackage{stfloats}
\usepackage{amsfonts}
\usepackage{bbm}
\usepackage{doi}
\usepackage{float}
\usepackage{amsmath,graphicx}
\usepackage{epsf}
\usepackage{epsfig}
\usepackage{epstopdf}
\usepackage{algorithm}
\usepackage{algorithmic}
\begin{document}
\title{Jamming in the Internet of Things: A Game-Theoretic Perspective}

\author{\IEEEauthorblockN{Nima Namvar$^\ast$, Walid Saad$^\dag$, Niloofar Bahadori$^\ast$, and Brian Kelley$^\ddag$ }
\IEEEauthorblockA{$^\ast$Department of Electrical and Computer Engineering, North Carolina A\&T State University, Greensboro, NC, USA \\
$^\dag$Wireless@VT, Bradley Department of Electrical and Computer Engineering, Virginia Tech, Blacksburg, Virginia, USA \\
$^\ddag$Department of Electrical and Computer Engineering, University of Texas at San Antonio, San Antonio, TX, USA\\
Emails: nnamvar@aggies.ncat.edu, walids@vt.edu, nbahador@aggies.ncat.edu,brian.kelley@utsa.edu}}
\IEEEtitleabstractindextext{
\begin{abstract}
Due to its scale and largely interconnected nature, the Internet of Things (IoT) will be vulnerable to a number of security threats that range from physical layer attacks to network layer attacks. In this paper, a novel anti-jamming strategy for OFDM-based IoT systems is proposed which enables an IoT controller to protect the IoT devices against a malicious radio jammer. The interaction between the controller node and the jammer is modeled as a Colonel Blotto game with continuous and asymmetric resources in which the IoT controller, acting as defender, seeks to thwart the jamming attack by distributing its power among the subcarries in a smart way to decrease the aggregate bit error rate (BER) caused by the jammer. The jammer, on the other hand, aims at disrupting the system performance by allocating jamming power to different frequency bands. To solve the game, an evolutionary algorithm is proposed which can find a mixed-strategy Nash equilibrium of the Blotto game. Simulation results show that the proposed algorithm enables the IoT controller to maintain the BER above an acceptable threshold, thereby preserving the IoT network performance in the presence of malicious jamming.
\end{abstract}

}

\maketitle

\IEEEdisplaynontitleabstractindextext
\IEEEpeerreviewmaketitle

\section{Introduction}
The Internet of Things (IoT) which will constitute a pervasive fabric that will interconnect a massive number of machines and human devices is one of the most critical technologies of the coming decade \cite{bahga2014internet}. The IoT will typically encompass a network of uniquely identifiable interconnected virtual and physical objects, such as sensors, radio frequency identification (RFID) tags, actuators, and mobile phones which are able to communicate and exchange data with each other to perform different tasks.

Given its large scale and the heterogeneity of its environment, the IoT is more vulnerable to security threats than other networks, such as cellular systems \cite{multimediaSecurity}. Thus, security requirements in the IoT will be more stringent than those of conventional wireless systems. Indeed, in the absence of robust security solutions, attacks and malfunctions in the IoT may outweigh any of its benefits. Due to its large attack surface, the security of an IoT system is prone to a variety of attacks such as malicious radio jamming, denial of service (DoS), side channel attack (SCA), replay attack, node capture, Sybil attack, and wormhole attack \cite{attackSurvey}.

In particular, IoT systems are vulnerable to jamming attack which could make the battery of target devices to drain quickly by disrupting their data transmission and making them retransmit repeatedly. Moreover, jamming could also lead to DoS which is the most common attack in wireless sensor networks (WSNs) and internet.

Several anti-jamming solutions for WSNs and other networks have been proposed in literature such as \cite{DoS,WSNjamming,OFDMjournal,OFDMciss}. However, most of the anti-jamming and jamming-detection techniques proposed for conventional communication networks are not appropriate for implementation in the IoT systems for two main reasons. First, IoT devices are resource constrained in terms of memory, processing capabilities, and transmission power and thus, are not capable of accommodating to complex or high power-consuming protocols \cite{resourceConstrainedIoT}. Moreover, distributed anti-jamming algorithms which rely on nodes' performance are not practical in the context of IoT systems in which the nodes are heterogeneous with diverse capabilities and demands. Therefore, devising anti-jamming protocols which can capture the specific requirements of IoT system is of crucial importance for the security of IoT devices. Nevertheless, only few studies have considered to address this problem.

The authors in \cite{Walid} studied a scenario in which the nodes in the IoT system possess different levels of importance based on their betweenness centrality. A jammer aims at disturbing an IoT network performance while remaining undetected by deliberately limiting its interference power and at the other hand, the IoT controller attempts at detecting the jamming attack by providing each node with some information bits proportionate to its centrality to announce its level of interference. In the case of successful detection, the nodes under attack increase their transmission power to overcome the jamming interference. In \cite{informationFusion}, the authors proposed a quarantine-based defense scenario in which the IoT devices report their level of interference to the IoT controller and on the interference level, the IoT controller may isolate the nodes under jamming attack to protect the rest of the network. Thus, to detect and/or confront the jamming attack, both studies \cite{Walid} and \cite{informationFusion} rely on some level of cooperation between the IoT devices and the IoT controller by using a distributed algorithm. However, implementing a distributed jamming detection scheme may leave the IoT system more vulnerable to targeted jamming attack when the IoT devices are resource constrainted and not capable of handling complex security mechanisms. Therefore, there is a need for an anti-jamming mechanism which can capture the specific requirements of IoT devices and accounts for their limited computational and energy resources.

The main contribution of this paper is to propose a centralized anti-jamming technique for OFDM-based IoT systems in which the IoT controller must defend against a jamming attack without relying on passive IoT nodes. The jammer attempts to compromise the network's functionality by allocating its power over different subcarriers (tones) of the system so as to maximize the number of affected devices. In our model, unlike the work in \cite{Walid}, the IoT nodes are assumed to be passive and resource limited and thus, they are not required to increase their transmit power to mitigate the jamming attack. Instead, the IoT controller responds to the jamming attack by intelligently allocating its power over the subcarriers in such a way that protects as many IoT nodes as possible. Moreover, to address the jamming attack, the IoT controller acts alone and does not require any cooperation from the resource-constrained IoT nodes. The problem is modeled as a Colonel Blotto game \cite{roberson2006colonel} with asymmetric resources between the IoT controller and the jammer and different scenarios based on the power budget of the IoT controller and the jammer are studied. To solve this game, we develop a novel evolutionary game-theoretic approach to find the Nash equilibrium (NE) strategies for the IoT controller and the jammer. Simulation results demonstrate the practicality of the proposed approach in terms of maintaining the bit error rate (BER) of the channels in an IoT system above an acceptable threshold.

The rest of the paper is organized as follows. In Section \ref{SEC:Model} the system model is presented and the problem is formulated as a zero-sum Blotto game. In Section \ref{SEC: Algorithm}, a novel algorithm based on evolutionary games is proposed to solve the game. Simulation results are provided in Section \ref{SEC: Simulations} and finally, conclusions are drawn in Section \ref{SEC: Conclusion}.

\section{System Model and Problem Formulation}\label{SEC:Model}
Consider an OFDM-based IoT system consisting of an access point (AP) and a set of $N$ IoT devices connected to the AP for communicating with one another and sharing data. The IoT devices are connected to the AP over an IEEE 802.11ah (low-power WiFi) protocol which is known to be a suitable wireless solution for resource-constraint IoT devices \cite{IoTconnectivity}. Let $\mathcal{D}=\{d_1, d_2,...,d_N\}$ and $\mathcal{S}=\{s_1, s_2,...,s_M\}$ be respectively, the set of $N$ IoT devices and the set of $M$ subcarriers. Given that the sub-carriers' bandwidth is relatively small compared to the total bandwidth of the system, the subcarriers are modeled as flat-fading Rayleigh channels between the devices and AP.

We consider a jammer $\mathfrak{J}$ which aims at disturbing the performance of IoT system by transmitting over specific subcarriers (tones), thereby interfering with the communication of IoT devices and increasing their BER. Indeed, the authors in \cite{BERofdm} show that the BER of OFDM systems, such as IEEE 802.11ah standard, is a function of the signal-to-interference-plus noise ratio (SINR) over all subcarriers that are allocated to a given user, irrespective of the baseband signal mapping scheme (e.g. QPSK or BPSK). Considering the fact that the jamming signal and the information signal are both independently attenuated by the channel, the instantaneous SINR over a subcarrier $s$ is given by:

\begin{equation}\label{SINR-Subcarrier}
  \gamma_s=\frac{g^{AP}_{s}p_{s}}{N_0+g^J_{s}j_{s}}, \forall s \in \mathcal{S}
\end{equation}
where $g^{AP}_{s}$ and $g^{J}_{s}$ are the channel power gains over subcarrier $s$ for the AP and the jammer $\mathfrak{J}$, respectively. Also, $p_s$, $j_s$, and $N_0$, denote the signal power over the subcarrier $s$, the jamming power over subcarrier $s$, and the noise power, respectively. The noise power is assumed to be constant over all subcarriers.



We assume that the access point AP and the jammer $\mathfrak{J}$ have a limited power budget, denoted by $P^{AP}$ and $P^{J}$, respectively. The jammer attempts at compromising the system performance by allocating its power over the subcarriers so as to degrade the received SINR and increase the BER. Suppose that, to successfully transmit the information packets over the subcarrier $s$, the received SINR $\gamma_s$ must exceed a threshold $\tau$. Therefore, the condition of having a successful transmission over subcarrier $s$ is given by,

\begin{equation}\label{SuccessfulTransmissionCondition}
  \frac{g^{AP}_{s}p_{s}}{N_0+g^J_{s}j_{s}}\geq\tau.
\end{equation}

\subsection{Problem Formulation}\label{subsec:Problem Formulation}
To thwart the jamming attack, the AP must distribute its power over the set of subcarriers in an optimal way which minimizes the effect of jamming and keeps the SINR $\gamma_s$ above $\tau$ for as many subcarriers $s \in \mathcal{S}$ as possible. However, the AP does not have any prior information about which subcarriers the jammer may decide to jam, and, thus, it has to act proactively based on the past observations of the jammer's behavior. Analogous to \cite{Walid}, we define the AP's payoff as the number of successful transmissions over $M$ available subcarriers,

\begin{equation}\label{utilityAP}
  U_{AP}(p_s, j_s)=\frac{1}{M}\parallel\{s\mid \gamma_s\geq\tau\}\parallel,
\end{equation}
where $\parallel.\parallel$ denotes cardinality of the set. Similarly, we also quantify the jammer's payoff as the number of unsuccessful transmissions over $M$ available subcarriers,
\begin{equation}\label{zeroSum}
  U_{J}(p_s, j_s)=\frac{1}{M}\parallel\{s\mid \gamma_s<\tau\}\parallel.
\end{equation}

The utility functions in (\ref{utilityAP}) and (\ref{zeroSum}) show that the payoff of each player depends on both its own power allocation strategy and that of its opponent. Therefore, we can model the interactions between the players in the context of game theory which is a suitable mathematical tool to model such resource allocation problems \cite{WalidToturial}. We model the problem in the framework of Colonel Blotto game \cite{roberson2006colonel}, which is a fundamental model of strategic resource allocation in multiple dimensions. In a Colonel Blotto game, two players compete over a number of independent battlefields by distributing their forces over these battlefields and within each battlefield, the player that allocates the higher level of force wins. Formally, the proposed Colonel Blotto game is defined as follows:

\begin{definition}
The \emph{Colonel Blotto} game which is labeled as $\mathbb{CB}\{AP,\mathfrak{J},M\}$ is a one-shot game between two players, namely AP and $\mathfrak{J}$, compete by simultaneously distributing their power over the set of $M$ subcarriers subject to their power budget constraints. Each battlefield, i.e. each subcarrier $s\in \mathcal{S}$, is won by the AP if transmission over $s$, as a function of power allocation strategy by both players, is successful and is won by the $\mathfrak{J}$, otherwise. The payoff for the whole game is the proportion of the wins on the individual battlefields.
\end{definition}

\begin{figure}
  \begin{center}
    \includegraphics[width=7cm]{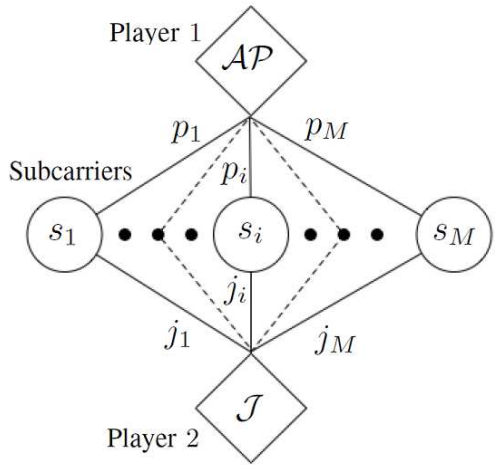}
   \caption{The Colonel Blotto game: AP and $\mathfrak{J}$ distribute their resources over the set of $M$ subcarriers.}\vspace{-.5cm}
   \label{fig:NetModel}
  \end{center}
\end{figure}

Therefore, we have a noncooperative zero-sum game for which the Nash equilibrium is considered as a suitable solution concept, where each player's payoff is maximized given the strategy of the opponent. However, the authors in \cite{roberson2006colonel} have shown that the Blotto game has no Nash equilibrium in pure strategies. Indeed, given any pure strategy vector, the opponent can always modify its resource allocation strategy accordingly to improve its payoff and win the game. Therefore, players tend to randomize their choice among several pure strategies, thereby employing a \emph{mixed strategy}.

A mixed strategy is a probability distribution over the set of all possible power-allocation vectors. Let $f(x_1, x_2, ..., x_M)$ and $h(x_1, x_2, ..., x_M)$ denote the probability distributions for the vector of resources $ \boldsymbol{x}=[x^j_1, x^j_2, ..., x^j_M]$ allocated to the subcarriers $\{s_i\}_{i=1}^{M}$ by AP and $\mathfrak{J}$, respectively. Then, the mixed strategy Nash equilibrium over the power allocation vector $\boldsymbol{x}$ is defined as,

\begin{align}\label{NashEquilibDefinition}
U_{AP}(f^*(\boldsymbol{x}),h^*(\boldsymbol{x})&\geq U_{AP}(f(\boldsymbol{x}),h^*(\boldsymbol{x})), \nonumber \\
U_{J}(f^*(\boldsymbol{x}),h^*(\boldsymbol{x}))&\geq U_{J}(f^*(\boldsymbol{x}),h(\boldsymbol{x})), \\
\forall f(\boldsymbol{x})^*\in & \Pi_{AP},h^*(\boldsymbol{x})\in \Pi_{J}, \nonumber
\end{align}
where $\Pi_{AP}$ and $\Pi_{J}$ denotes the set of all possible mixed strategies of the AP and the $\mathfrak{J}$, respectively.

Note that because of the a priori equivalence of all subcarriers, the marginal distributions of the mixed Nash equilibrium must be identical, i.e. $f(x_1)=f(x_2)=...=f(x_N)=f(x)$ and $h(x_1)=h(x_2)=...=h(x_N)=h(x)$. Therefore, the optimization can be carried out over the marginal distributions $f(x)$ and $h(x)$ \cite{roberson2006colonel}. Let $F(x)$ and $H(x)$ denote the marginal cumulative distribution functions (CDFs) of AP and $\mathfrak{J}$, respectively. In our game, the AP seeks to maximize its expected payoff subject to its power budget. Therefore, the optimization problem for the AP will be:

\begin{align}\label{APoptimization}
& \underset{F(p_s)}{\text{maximize}}
& & \frac{1}{M}\sum_{s=1}^{M}\int_{0}^{\infty}\text{Pr} \left( p_s\geq \tau(\frac{N_0+g_s^{J}j_s}{g_s^{AP}})\right)dF(p_s) \\
& \text{s.t.}
& & \sum_{s=1}^{M}\int_{0}^{\infty}p_sdF(p_s)=P^{AP}, \; s = 1, \ldots, M.
\end{align}
where (6) shows the expected payoff of AP and (7) ensures that the total power of AP is restricted to its power budget $P^{AP}$. Similarly, the jammer seeks to maximize its expected payoff according to the following optimization problem:

\begin{align}\label{Joptimization}
& \underset{H(j_s)}{\text{maximize}}
& & \frac{1}{M}\sum_{s=1}^{M}\int_{0}^{\infty}\text{Pr} \left(\frac{1}{g^J_s}(\frac{g^{AP}_sp_s}{\tau}-N_0) \right)dH(j_s) \\
& \text{s.t.}
& & \sum_{s=1}^{M}\int_{0}^{\infty}j_sdH(j_s)=P^{J}, \; s = 1, \ldots, M.
\end{align}
where (8) shows the expected payoff of the jammer and (9) ensures that the total power of jammer is restricted to its power budget $P^{J}$.

Note that the optimization problems in (\ref{APoptimization}) and (\ref{Joptimization}) are coupled in the sense that any choice of $H(j_s)$ in (\ref{Joptimization}) will affect the optimal solution $F(p_s)$ in (\ref{APoptimization}) and vice versa. Therefore, we can conclude that the solution of (\ref{APoptimization}) is a probability distribution $F(p_s)$ over the set of subcarriers which results in the maximum AP's payoff for a \emph{given} strategy of the jammer. A similar argument is applicable to the solution of (\ref{Joptimization}). Hence, the solution of the above coupled optimization problems, $(F^*(p_s), H^*(j_s))$ is a pair of probability distribution functions which are the best response to one another. In other word, if the AP chooses the strategy $F^*(p_s)$, the jammer cannot do better than selecting $H^*(j_s)$ as its own strategy. Therefore, $(F^*(p_s), H^*(j_s))$ is the Nash equilibrium of the proposed Blotto game.


The solution for Blotto games with deterministic payoff functions can be found in \cite{roberson2006colonel,OFDMjournal}. However, obtaining the closed-form solutions to the optimization problems in (\ref{APoptimization}) and (\ref{Joptimization}) is particularly challenging due the presence of random channel gains $g_s^{J}$ and $g_s^{AP}$ in the payoff functions. Nevertheless, we can acquire some insights about the set of possible mixed strategy solutions by closely looking at the problem. First, we note that, assuming $g_s^{AP}=g_s^{J}=1$, if $P^{AP}>M(P^{J}+N_0)$, the game has a set of trivial optimal pure strategies for AP, with $p_s=\frac{P^{AP}}{M}$, for all $s=1,2,...,N$, which results in a payoff of $1$ for the AP and a payoff of $0$ for the jammer. In other words, the AP has enough power to keep the subcarriers' SINR above $\tau$ irrespective of power allocation strategy of jammer. Furthermore, we note that in the equilibrium strategy, the power allocated to each subcarrier by the AP is lower bounded by $\frac{\tau N_0}{g_s^{AP}}$, because if the AP allocates power in the range $[0, \frac{\tau N_0}{g_s^{AP}}]$, the transmission will always fail in subcarrier $s$. On the other hand, $\mathfrak{J}$'s allocation is lower bounded by $0$ because it does not have any obligation to distribute its power over \emph{all} subcarriers.


\section{Proposed Evolutionary Algorithm}\label{SEC: Algorithm}
To find the best strategies in the anti-jamming problem, we adopt an evolutionary-based algorithm \cite{evolutionary}. Consider a population of size $K=K_{AP}+K_{J}$ of $K_{AP}$ players of the type AP and $K_{J}$ players of the type $\mathfrak{J}$. Players of the same type have identical resources; however, each player may adopt a different strategy profile. The strategy of each player $i$ at time step $t$ is characterized by its probability distribution $f_i(x, t)$ or $h_i(x,t)$ if $i$ is of type AP or of type $\mathfrak{J}$, respectively. $f_i(x, t)$ and $h_i(x,t)$  are probabilities of allocating $x$ resources at any given subcarrier at time $t$ by player $i$. The Blotto game is played successively between all pairs of opponents and, as explained below, the  players' strategies evolve in response to the resulting payoffs. At each time step, each player of type AP faces the opponents of the opposite type and its expected payoff function is given by,

\begin{algorithm}[t]
\begin{algorithmic}
\STATE \textbf{Initialization}: Each player $i \in \{1,2,...,K\}$ starts with an initial probability distribution
\REPEAT
    \FORALL {player $i \in \{1,2,...,K\}$}
        \STATE compute the expected payoff according to ((\ref{ExpectedPayoffAP}) and (\ref{ExpectedPayoffJ}))
            \FORALL {player $i \in \{1,2,...,K\}$}
                \STATE (I)select another player $i'$ of the same type as $i$
                    \IF {$\pi_i<\pi_{i'}$}
                        \STATE $i$ adopts the strategy of $i'$
                        \COMMENT{\emph{Imitation}}
                    \ENDIF
                    \STATE select a random number $\beta$ in range $[0,1]$
                    \IF {$\beta<\xi$}
                        \STATE Subtract small amount $2q$ from a randomly selected point in strategy vector and add $q$ to two other random points
                        \COMMENT{\emph{Mutation}}
                    \ENDIF

            \ENDFOR

    \ENDFOR
\UNTIL {convergence}

\end{algorithmic}
\caption{Proposed algorithm for solving the resource allocation problem}
\end{algorithm}

\begin{table}\label{tab:parameters}
\centering
\caption{Simulation Parameters}
    \begin{tabular}{lc}
    \hline
        
        Parameter &     Value \\ \hline
        $K_{AP}$        & $100$     \\
        $K_{J}$         & $100$     \\
        $\xi$           & $1.25\times 10^{-3}$     \\
        $P^{AP}$        & $1 W$    \\
        $P^{J}$         & $100 mW$     \\
        $M$             & $128$     \\
        $N_0$           & $-90 dBm$     \\
        $\tau$          & $25 dB$     \\ \hline
    \end{tabular}
\end{table}

\begin{align}\label{ExpectedPayoffAP}
\pi_{AP}^i= & \frac{1}{K_{J}}\sum_{z=1}^{K_{J}}\int_{0}^{P^{AP}}\int_{0}^{P^{J}}U_{AP}(f_i(x,t),h_z(x',t))dxdx', \\
\forall i &\in \{1, 2, ..., K_{AP} \}   \nonumber
\end{align}
Similarly, at each time step, each player of type $\mathcal{J}$ faces the opponents of the opposite type and the its expected payoff function is given by,
\begin{align}\label{ExpectedPayoffJ}
\pi_{J}^i= &\frac{1}{K_{AP}}\sum_{z=1}^{K_{AP}}\int_{0}^{P^{J}}\int_{0}^{P^{AP}}U_{J}(f_z(x,t),h_i(x',t))dxdx', \\
\forall i &\in \{1, 2, ..., K_J \} \nonumber
\end{align}

The strategies of the players evolves according to their expected payoff at each time step $t$. The evolution of strategies is based on the following  dynamical rules:

\begin{itemize}
  \item \emph{Imitation}: At each time step $t$, for each player $i$, another player $i'$ of the same kind is randomly selected from the entire population and their average payoffs are compared. If $\pi_i<\pi_{i'}$, player $i$ \emph{imitates} player $i'$ by adopting its strategy $f_{i'}(x)$ or $h_{i'}(x)$, otherwise, the strategy of player $i$ does not change.
  \item \emph{Mutation}: To increase the diversity of the strategy sets, we assume that the strategy of each player $i$ is subject to a process of mutation with small probability $\xi$ in which, an amount of $2q$ is subtracted from the strategy of $i$ at a randomly selected point and the amount $q$ is added to the two nearest points in the mixed strategy vector.
\end{itemize}

In order to implement the mentioned dynamical rules numerically, we discretize the resource variable $x$ to a vector of $L$ equally distant points in the feasible intervals $[0, P^{AP}]$ and $[0, P^{J}]$ for the type AP and the type $\mathfrak{J}$, respectively.

Note that in each iteration, each player's strategy evolve in response to the opponent's strategy to maximize its payoff, according to the coupled optimization problems in (\ref{APoptimization}) and (\ref{Joptimization}). Hence, upon convergence, the proposed evolutionary algorithm finds the best strategies of the players which are the Nash equilibrium of the proposed Blotto game according to the previous discission. Also, note that the imitation process requires the players to constantly adopt the strategy of a better player in each iteration. Meanwhile, the mutation process, enables exploring a diversity in the set of strategies in each round, which can help to find the best strategies for both players.

\section{Simulations Results}\label{SEC: Simulations}
For our simulations, we consider an IoT access point with power budget $1$~W which distribute its power over $128$ OFDM subcarriers to confront a jammer with power $50$~mW. The SINR threshold for successful transmission over each subcarrier is $25$~dB. For simplicity, we assume that the noise power is constant over all channels of the system. Other simulation parameters are listed in Table I.

\begin{figure}[t]
  \begin{center}
    \includegraphics[width=8cm]{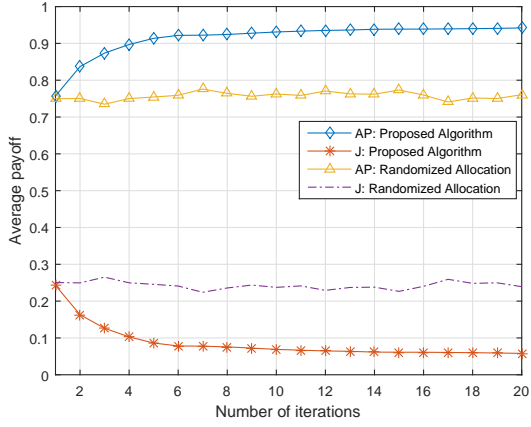}
   \caption{The average payoff of AP and $\mathfrak{J}$ according to the proposed algorithm and random allocation strategy.}\vspace{-.5cm}
   \label{sim1}
  \end{center}
\end{figure}

Figure \ref{sim1} shows the expected payoff of the access point and the jammer as the players' strategies evolve according to the proposed algorithm. In Figure \ref{sim1} we can see that, as time evolves, the access point updates its power allocation strategy and can effectively mitigate the jamming attack and securing a reliable communication over $96\%$ of time for each subcarrier. For comparison, in Figure \ref{sim1}, we also show the average payoff of the players according to randomized allocation. Clearly, due to the lack of \emph{learning} and \emph{adaptation}, the randomized allocation approach can save only $75\%$ of the channels, leading to a poor system performance in the presence of jamming attack. Figure \ref{sim1} shows that the proposed algorithm achieves up to $30.67\%$ gain over the random power allocation approach.

\begin{figure}[t]
  \begin{center}
    \includegraphics[width=8cm]{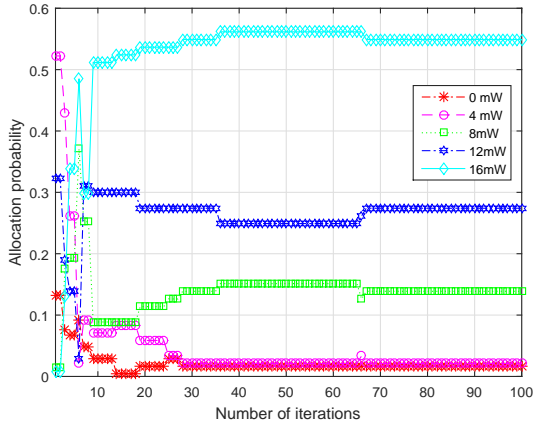}
   \caption{The evolution of AP's mixed strategy. The power of AP for each subcarrier is discretized to $L=5$ levels to show the convergence of the proposed algorithm.}\vspace{-.5cm}
   \label{sim2}
  \end{center}
\end{figure}
\begin{figure}[t]
  \begin{center}
    \includegraphics[width=8cm]{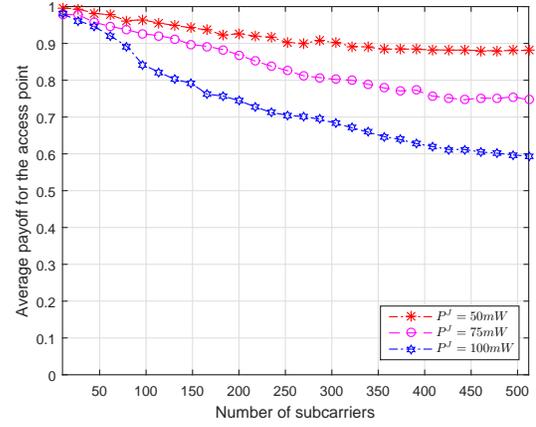}
   \caption{Average defender's payoff for different network sizes.}\vspace{-.5cm}
   \label{sim3}
  \end{center}
\end{figure}

Figure \ref{sim2} shows the evolution of the mixed strategy for the AP. For illustration simplicity, we assume that the AP's power vector for each subcarrier is discretized to $L=5$ levels in the set $\{0, 4, 8, 12, 16\}$~mW. In this figure we can see that the proposed algorithm converges after a reasonable number of $20$ iterations.

Figure \ref{sim3} shows the IoT defender's performance as a function of network size for $3$ different jamming power. As the number of subcarriers increases, the access point has to allocate its power over a larger set of channels. As previously discussed, the power allocated to each subcarrier by the IoT access point is lower-bounded by $\frac{\tau N_0}{g_s^{AP}}$. Therefore, as the network's size increases, the IoT access point has less power to distribute among the subcarriers in order to combat the jamming, since it is required to provide the minimum amount of power, $\frac{\tau N_0}{g_s^{AP}}$, for all subcarriers. Figure \ref{sim3} also demonstrates the effect of the jamming power budget on the system's performance. As the power of jammer increases, the utility of access point decreases which means that a higher percentage of channels suffer a low SINR due to the jamming attack. However, as seen in Figure \ref{sim3}, the proposed approach is able to protect up to $95\%$ and $85\%$ of channels in the presence of jammer with power budget $50$~mW for a network size of $128$ and $256$ subcarriers, respectively.

\begin{figure}[t]
  \begin{center}
    \includegraphics[width=8cm]{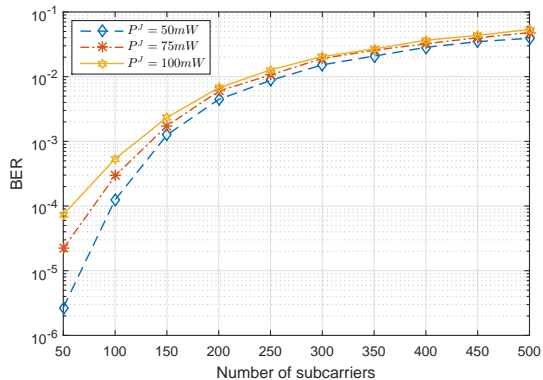}
   \caption{Average BER for different network sizes.}\vspace{-.7cm}
   \label{sim4}
  \end{center}
\end{figure}

Figure \ref{sim4} shows the average BER of the sytem as a function of the network size. In this figure, we can see that, by increasing the network size, the BER increases since by distributing the limited power of access point to a larger set of channels, the SINR of each channel will decrease. Figure \ref{sim4} also shows that for small network sizes, the jamming power is the dominant factor in determining the average BER of each channel. However, as the number of subcarriers increases beyond $350$, the jamming power has less effect on the BER.

\begin{figure}[t]
  \begin{center}
    \includegraphics[width=8cm]{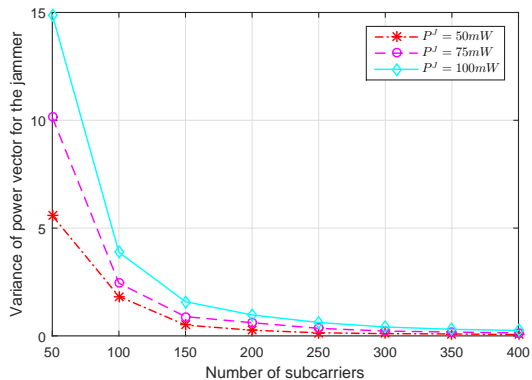}
   \caption{Variance of the jammer's power allocation vector for different network sizes.}\vspace{-.7cm}
   \label{sim5}
  \end{center}
\end{figure}

Figure \ref{sim5} shows the variance of the jammer's power vector $[j_1, j_2, \dots, j_M]$ as a function of network size. It can be seen from Figure \ref{sim5} that, for small network sizes, when the AP has enough power to allocate to all subcarriers, the jammer adopts a high variance strategy. Under such strategy, the jammer allocates either very small or very large portion of its budget to different subcarriers, so as to compromise the performance on a subset of those subcarriers. However, as the network size increases and the AP's power is allocated over a larger number of subcarriers, the jammer tends to select a low variance strategy under which, it can compromise more subcarriers by allocating more or less equal portion of its budget to a subset of subcarriers, as the AP does not have enough resources to defend all subcarriers.

\section{Conclusion}\label{SEC: Conclusion}
In this paper, we have studied the problem of jamming in an IoT system. In particular, we have proposed a centralized mechanism to address the jamming problem in an IoT system composed of resource constrained devices. In the proposed model, an IoT access point defends against the jammer by allocating its power over the subcarriers in a smart way to defend the system. We have modeled the problem as a Colonel Blotto game between the IoT access point and the jammer. In this game, each player seeks to maximize its payoff by adopting a randomized strategy to allocate its power to the subcarriers. To solve this game, we have proposed a novel evolutionary-based algorithm to find the equilibrium strategies for both players. In the proposed algorithm, the players adapt their strategies according to some dynamical rules to maximize their payoff given the opponent's strategy. Simulation results have shown that the proposed algorithm converges in a reasonable number of iterations and yields considerable performance advantage compared to the random power allocation strategy.

\section{Acknowledgement}
The authors would like to acknowledge the support from Air Force Research Laboratory and OSD for sponsoring this research under agreement number FA8750-15-2-0116. The U.S. Government is authorized to reproduce and distribute reprints for Governmental purposes notwithstanding any copyright notation thereon. The views and conclusions contained herein are those of the authors and should not be interpreted as necessarily representing the official policies or endorsements, either expressed or implied, of Air Force Research Laboratory, OSD, or the U.S. Government. This work was also supported,
in part, by the U.S. National Science Foundation under Grants
CNS-1524634 and ACI-1541105.

\bibliography{myref}
\bibliographystyle{IEEEtran}

\end{document}